\documentclass[%
aps,pre,%
 amsmath,amssymb,
reprint,%
superscriptaddress,showpacs]{revtex4-1}

\usepackage{color, graphicx}
\usepackage{dcolumn}
\usepackage{bm}
\usepackage{amssymb}
\usepackage{latexsym}
\usepackage{amsfonts}
\usepackage{mathtools} 
\usepackage{epstopdf}  
\usepackage{amsmath}
\usepackage{multirow}
\usepackage{ifthen}
\usepackage{float}
\usepackage{lipsum}

\begin{document}

\title{Wave propagation in a strongly disordered 1D phononic lattice supporting rotational waves}
\author{A.~Ngapasare}
\author{G.~Theocharis}
\author{O.~Richoux}
\affiliation{ Laboratoire d'Acoustique de l'Universit\'e du Mans, UMR CNRS 6613 Av. O. Messiaen, F-72085 LE MANS Cedex 9, France}
\author{Ch.~Skokos}
\affiliation{Department of Mathematics and Applied Mathematics, University of Cape Town, Rondebosch 7701, South Africa}
\author{V.~Achilleos}
\affiliation{ Laboratoire d'Acoustique de l'Universit\'e du Mans, UMR CNRS 6613 Av. O. Messiaen, F-72085 LE MANS Cedex 9, France}

\begin{abstract}
We investigate the dynamical properties of a strongly disordered micropolar lattice
made up of cubic block units. This phononic lattice model supports both transverse
and rotational degrees of freedom hence its disordered variant posses an interesting problem as 
it can be used to model physically important systems like beam-like microstructures. Different kinds of single site excitations (momentum or displacement) on the two degrees of freedom
are found to lead to different energy transport both superdiffusive and subdiffusive.
We show that the energy spreading is facilitated both by the low frequency extended waves and 
a set of high frequency modes located at the edge of the upper branch of the periodic case
for any initial condition.
However, the second moment of the energy distribution strongly depends on the initial condition and it is slower
than the underlying one dimensional harmonic lattice (with one degree of freedom). Finally, a limiting case of the micropolar
lattice is studied where Anderson localization is found to persist and no energy spreading
takes place.
\end{abstract}

\maketitle

\section{Introduction} \label{sec1}
Wave propagation in heterogeneous media has attracted tremendous research interest in recent years. 
Families of one-dimensional ($1$D) continuous and discrete models have been extensively studied in this context~\cite{fpu,disorder_14,izrailev,disorder_15}
focusing on the localization properties of both the normal modes of finite systems i.e.,~Anderson localization (AL)~\cite{anderson}, and on the wave propagation in infinite media. 
Although the theory of AL was initially formulated for electronic systems, it has been successfully extended and applied to many other systems.
Interestingly, recent experimental results on AL (see e.g.,~Refs.~\cite{billy,roati,schwartz,lahini,jk}) have opened new research frontiers and have revitalized the interest on studying AL both in quantum and classical systems. 
 
In the context of linear disordered $1$D lattices, among different systems, special attention has been given to the tight binding electron model,
the linear Klein-Gordon (KG) lattice~\cite{Ishii,izrailev} and the harmonic lattice~\cite{Kundu,Lepri}. The interest in these models lies partly in the fact that they represent 
the linear limit of seminal nonlinear lattices such as the discrete nonlinear Schr\"{o}dinger equation (DNLS), the quartic KG, and the  Fermi-Pasta-Ulam-Tsingou (FPUT) lattices~\cite{kevre,fpu,flach}. 
Even more, these fundamental models have been adopted to describe a variety of physical systems and more recently, in the context of metamaterials, they have been extensively used as toy models for novel wave phenomena~\cite{deymer2,flach}.

A typical route to study the wave properties of these heterogenous lattices is to monitor the time evolution of initially compact wave-packets. For the tight binding and the linear KG models, the dynamics after the excitation of such an initial condition is characterized by an initial phase of spreading, followed by a phase of total confinement to its localization length/volume. The width of the wave-packet is of the order of the maximum localization length~\cite{loc_len}. On the other hand, for the harmonic lattice, along with the localized portion of the energy, there is always a propagating part due to the existence of extended modes at low frequencies. A quantitative description of wave propagation in disordered $1$D systems of one degree of freedom (DOF) per lattice site was formulated in Refs.~\cite{Ishii,Kundu,Lepri} where wave-packet spreading was quantified using both analytical and numerical methods. Moreover, many variations of these $1$D lattices have been studied extensively in several works including all the regimes from the periodic linear to the disordered nonlinear~\cite{vassospre2016,jk,Chiaradis,Guebelle,Sen2017poly,MasonPanos,Luding2,arn_gran,many2}. 

A natural extension to the above studies is to investigate the corresponding behavior in disordered lattices with more than one DOF.
Few such studies already exist in the literature especially as generalizations of the tight binding model by assuming a linear coupling between two (or more) $1$D chains~\cite{ladderPRB,flachlieb} and illustrate how the coupling modifies the energy transport properties. Recent experiments also revealed the role of additional forces in disorder mechanical lattices.~\cite{beamexper}. On the other hand, the wave dynamics of disordered harmonic chains with two DOFs per site has merely been studied. Such models are relevant to macroscopic mechanical lattices (e.g.,~granular phononic crystals, lego and origami chains~\cite{pichard,florian,jk2,lego,Flornew}), where the coupling between the DOFs stems from either the geometrical characteristics or from the material properties.

Here, we present a thorough numerical study of a linear disordered system made up of square block elements that supports both translational and rotational waves~\cite{Vasiliev,Deymier,micromechanics}. The model we investigate is used in bodies with beam-like microstructure to construct continuum models and in beam lattices~\cite{Vasiliev,Noor}. The corresponding equations of motion bare close resemblance to other structures including $1$D lattices of elastic cylinders~\cite{pichard} or spherical beads~\cite{florian}. Our goal is to unveil the role of the coupling between the DOFs regarding the energy transport in the presence of strong disorder and to identify the differences with the underlying $1$D harmonic lattice.

The rest of this paper is arranged as follows: In Section~\ref{sec2} we describe the model supporting both transverse and rotational motion. The static properties for the periodic and disorder cases are also discussed. In section~\ref{sec3} we investigate in detail the dynamical behavior of the system in the presence of strong disorder by initially exciting a single DOF at the center of the lattice. In section~\ref{sec4} we summarize and conclude the paper.

\begin{figure} 
\includegraphics[width=7.50cm]{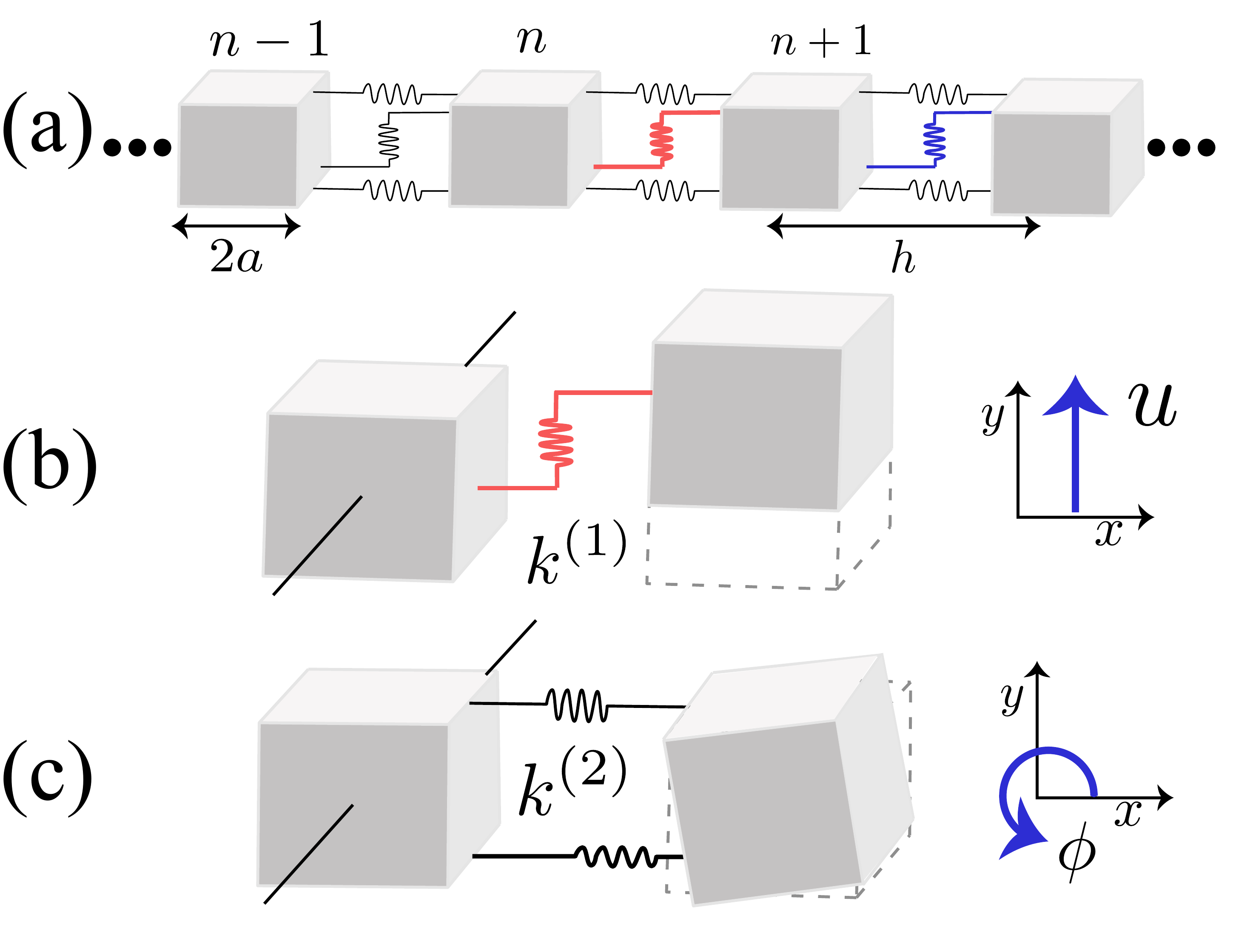}
\caption{(a) Schematic of the disorder phononic lattice with random shear stiffness indicated by the different spring thicknesses (colors). (b) Illustration of the transverse motion and the corresponding shear stiffness  $k^{(1)}$. (c) Illustration of rotational motion and the corresponding bending stiffness $k^{(2) }$.}
\label{model1}
\end{figure}

\section{ discrete model} \label{sec2}
We consider a phononic structure composed of discrete block-spring elements such that the $n$th element can be described by transverse and rotational DOFs as shown in Fig.~\ref{model1}. 
The transverse displacements are in the $y$ direction whilst the rotation is about an axis perpendicular to the $xy$-plane. The blocks are coupled through a shear stiffness $k^{(1)}_n$ and a bending one $k^{(2)}_n$ [see Fig.\ref{model1}(b,c)]. 
In this work we consider $N$ identical cube blocks of mass $m$ with edges of length $2a$ and consequently a moment of inertia $I=2ma^2/3$.  Systems that could be potentially described by such a structure include models in micro- and nano-scale films~\cite{Randow}, granular media~\cite{Flornew}, modeling of beam lattices~\cite{Noor} or the interaction of finite size particles with pre-designed connectors~\cite{Katia}.
The periodicity of the system is imposed by the distance $h$ between the center of each block as shown in Fig.~\ref{model1}, where $u_n$ and $\phi_n$ respectively represent the transverse and rotational motion of the $n$th block from equilibrium. The corresponding momenta are written as ${P_{n}^{(u)}}=m\dot{u}_n$ and ${P_{n}^{(\phi)}}=I\dot {\phi}_n$ for the former and latter motions, while $(~\dot{}~)$ denotes derivative with respect to time.
The total energy of the system $H$, of the system is given by the following expression~\cite{Deymier,Suiker}
\begin{widetext}
\begin{equation}
H  =\sum_{n=1}^{N}   \frac{ 1 }{2} {P_{n}^{(u)}}^{2}  +\frac{ 1 }{2I} {P_{n}^{(\phi)}}^{2}  
+   \frac{ 1 }{2}  K^{(1)}_{n+1} \Big[  (u_{n+1} - u_{n} )  +  \frac{3}{2} (\phi_{n+1} + \phi_{n} ) \Big] ^2 
+   \frac{ 1 }{2} K^{(2)}_{n+1} (\phi_{n+1} - \phi_{n} ) ^2.
\label{hamilt}
\end{equation}
\end{widetext}
Here we have defined the constants $ K^{(1)}_n = 2 k^{(1)}_n (2a)^2/l^4_d  $, $ K_n^{(2)} =  k^{(2)}_n2a^2/l^2  $, the lengths $ l =h -2a$ and $ l_d = \sqrt{l^2 + (2a)^2} $ and for simplicity we choose $ m =1 $,  $l=1$ and thus $h=3$. The equations of motion for the two DOF are explicitly given by:
\begin{align}  
  \ddot{u}_{n} & =  K^{(1)}_{n+1} \big( u_{n+1} - u_{n}\big) -  K^{(1)}_{n} \big(u_{n}   - u_{n-1}\big)\nonumber \\
  &+ \frac{ 3 K^{(1)}_{n+1}}{2} \big(  \phi _{n+1} +\phi _{n}  \big) -  \frac{3 K^{(1)}_{n} }{2}  \big( \phi _{n} +\phi _{n-1}  \big) ,\label{eqom1}
 \\
I\ddot{\phi}_{n}  & =   \frac{3 K^{(1)}_{n} }{2 } \big( u_{n-1} -u_{n} \big) +  \frac{3 K^{(1)}_{n+1}}{2} \big( u_{n} -u_{n+1} \big)\nonumber \\
 &-  \frac{9 K^{(1)}_{n+1}}{4} \big( \phi_{n+1}  + \phi_{n}\big)  -   \frac{9 K^{(1)}_{n} }{4}\big(\phi_{n} + \phi_{n-1}\big) \nonumber \\
  & + K^{(2)}_{n+1} \big( \phi_{n+1} -  \phi_{n}\big)  - K^{(2)}_{n} \big(\phi_{n} - \phi_{n-1} \big).
\label{eqom2}
\end{align}
We first study the periodic phononic crystal~\cite{pichard} with $K^{(1)}_n\equiv K^{(1)}=1$ and $K^{(2)}_n\equiv K^{(2)}$.
In this case, we may look for Bloch like solutions of the form
\begin{equation}  \label{eq3}
\mathbf{X}_n =  \begin{pmatrix} 
            u_n (t)\\
            \phi_n (t)
           \end{pmatrix} = \mathbf{X}e ^{i\Omega t - i Qn},
\end{equation}
 where $\mathbf{X}=[U,\Phi]$ is the amplitude vector, $\Omega$ is the frequency
and $ Q $ is the Bloch wave number.

\begin{figure} 
\includegraphics[width=8.1cm]{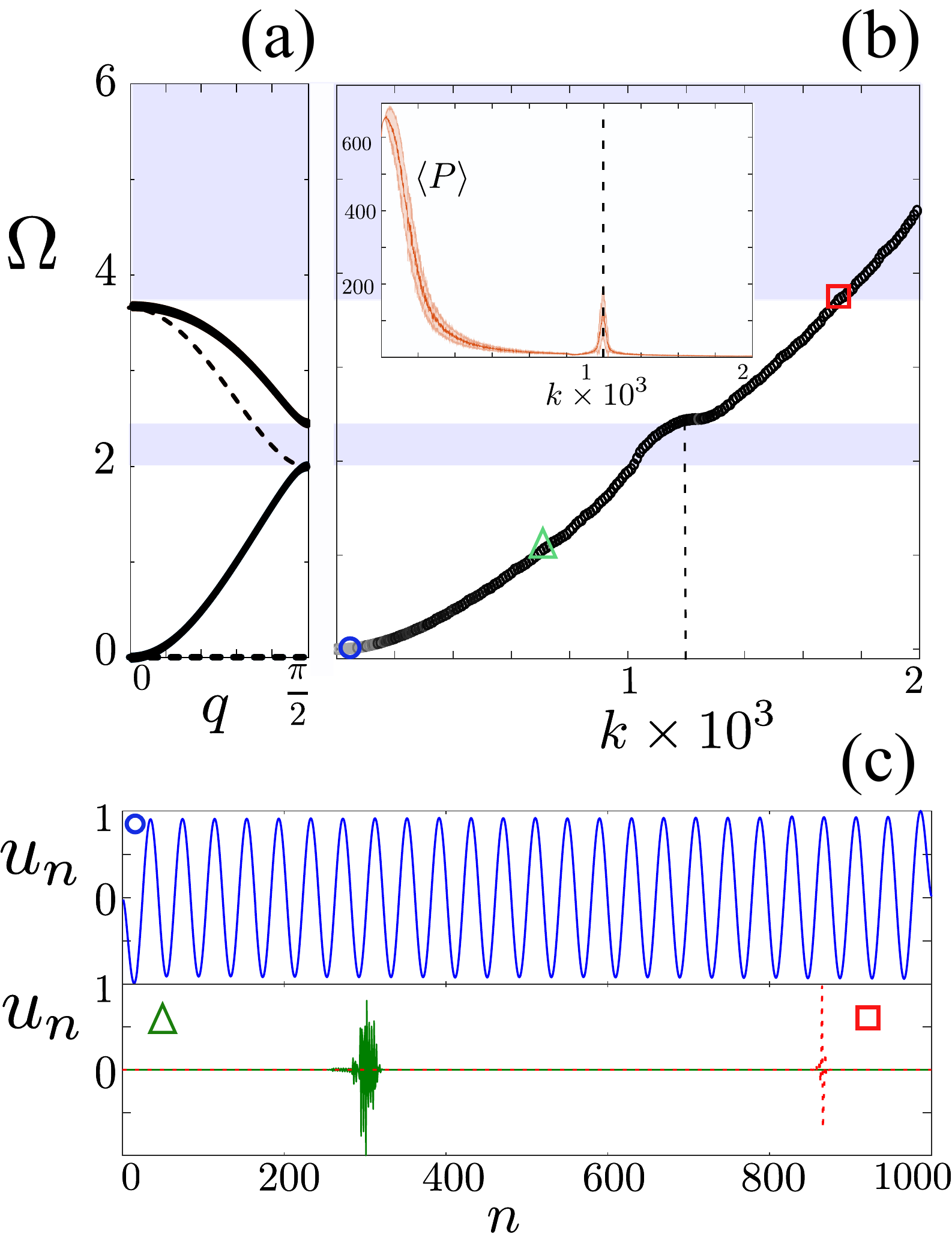}
\caption{ (a) Dispersion relations of the lattice for $K^{(1)}=1$ and $K^{(2)} = 1$ ($K^{(2)} = 0$) solid curves (dashed curves). (b) Corresponding eigenfrequencies for a single  strongly disordered lattice $(W=2)$ with $\langle K^{(1)}\rangle=1$ and $K^{(2)} = 1$. The inset shows the mean value ($200$ realizations) of $\langle P \rangle $ for each mode, and the standard deviation (shaded area). The vertical dashed line denotes the index where the \textit{quasi-extended} modes appear. (c) Representative profiles of the eigenmodes of the disordered lattice with $K^{(2)} = 1$ for the three different cases indicated by the circle, square and triangle in (b). Here we show only profiles for $u_n$. }
\label{profile}
\end{figure}

Inserting Eq.~(\ref{eq3}) into Eqs.~(\ref{eqom1}) and ~(\ref{eqom2}) we obtain the following eigenvalue problem for the allowed frequencies 
$\mathbf{S} \mathbf{X} = \Omega^2 \mathbf{X}$,
where the resultant dynamical matrix is
$$\mathbf{S} = \begin{pmatrix}
     4 \sin^2 q  & -6~i \sin q  \cos q \\
      6~i  \sin q  \cos q   & \frac{2}{3} [ 9\cos^2 q + 4 K^{(2)} \sin^2 q ] \\
         \end{pmatrix},
         $$
with $q = Q / 2 $. 
The corresponding expression for the eigenfrequencies is given by 
\begin{widetext}
\begin{equation}
\Omega^2_{\pm} = \frac{1}{2}    \Bigg\{  4 \sin^2 q +\frac{2}{3} \left( \frac{9}{4} 4 \cos^2 q  + 4 K^{(2)} \sin^2 q \right) \pm \sqrt{ \bigg [ 4 \sin^2 q + \frac{2}{3}\left( \frac{9}{4} 4 \cos^2 q  + 4 K^{(2)} \sin^2 q \right) \bigg ]^2  - 64 K^{(2)} p \sin^4 q } \Bigg \} .  \label{dispr}
\end{equation}
\end{widetext}
The dispersion relation of Eq.~(\ref{dispr}) for $K^{(2)}=1$ is depicted by the solid curves plotted in Fig.~\ref{profile}(a). 
We directly observe the appearance of two branches separated by a band gap and terminated by a maximum allowed frequency. Since the two DOFs are coupled, the modes are composed by a mixture of transverse and rotational motion. Note that the constant $K^{(2)}$, which depends on the bending stiffness, can be used as a tuning parameter to change the form of the dispersion relation and the dominant motion participating in each propagating mode~\cite{pichard}.

In the rest of this work we introduce disorder to the system only through the shear spring stiffness's $K^{(1)}_n$ [see also Fig.~\ref{model1}(a)].
We choose this particular disorder aiming to expose the role of each DOF and isolate its importance in
the energy transfer.
The values of  $K^{(1)}_n$ are taken from a uniform probability distribution 
\[
  f\bm{\big(} K^{(1)}_n \bm{ \big) } =
  \begin{cases}
                                   W^{-1}, &-W/2    < \text{ $K^{(1)}_n$} - \langle K^{(1)}\rangle < W/2, \\
                                   0 & \text{otherwise}. \\
    \end{cases}
\]

The parameter $W$ determines the width of the distribution and thus the strength of the disorder.
Fig.~\ref{profile}(b) illustrates the eigenfrequencies of a strongly disordered ($W=2$) finite chain of $N=10^3$ blocks.  
The eigenmodes have been sorted from lowest to highest frequency for increasing mode index $k$.
Due to the strength of the disorder, the middle band gap is filled with modes while the maximum frequency of the system is much bigger in comparison to the maximum frequency of the periodic chain.

To further characterize the disordered finite lattice, for each mode we calculate the participation number~\cite{flachbook} $ P = 1/ \sum h_n^2 $ where $h_n = H_n / H$ is the normalization of the site energy $H_n$. $P$ is an indicator of the localization of the mode and it becomes $P\approx N$ for a mode with almost all sites excited, while $P = 1$ for a single site mode. The mean value of $P$ taken for $200$ disorder realizations is shown in the inset of Fig.~\ref{profile}(b). It becomes clear that most of the modes are strongly localized throughout the spectrum except at very low frequencies where a rather small portion of the modes is extended.  As such, we may loosely describe the modes as either localized or extended. Interestingly enough we obtain a set of what we coin as \textit{quasi-extended} modes around the cut-off frequency of the upper branch of the periodic case. The appearance of these modes is due to the particular implemented disorder, which is only on the shear stiffness (see Appendix~\ref{sec6}). For illustration in Fig.~\ref{profile}(c) we show the normalized transverse profiles of an extended mode [$k= 50$ (circle)] and of two localized modes [$k=735$ (triangle), and $k=1700$ (square)]. The normalized rotational profile follows the same patterns. As we will show below, both the low frequency \textit{extended} modes and the \textit{quasi-extended} modes contribute to the transfer of energy in the lattice.

\section{Dynamics of the system} \label{sec3}

To study the properties of energy transfer in the system we excite strongly disordered lattices using single site initial conditions.
Our results are averaged over an ensemble of $200$ disorder realizations and since we are interested in the effects of strong disorder we choose $W = 2$ such that $ K^{(1)}_n \in (0, 2) $. For all the time dependent simulations, each realization had $N=10^5$ lattice sites.

\subsection{Momentum excitation} \label{ssec1}
\begin{figure}[ht]
\includegraphics[width=8.4cm]{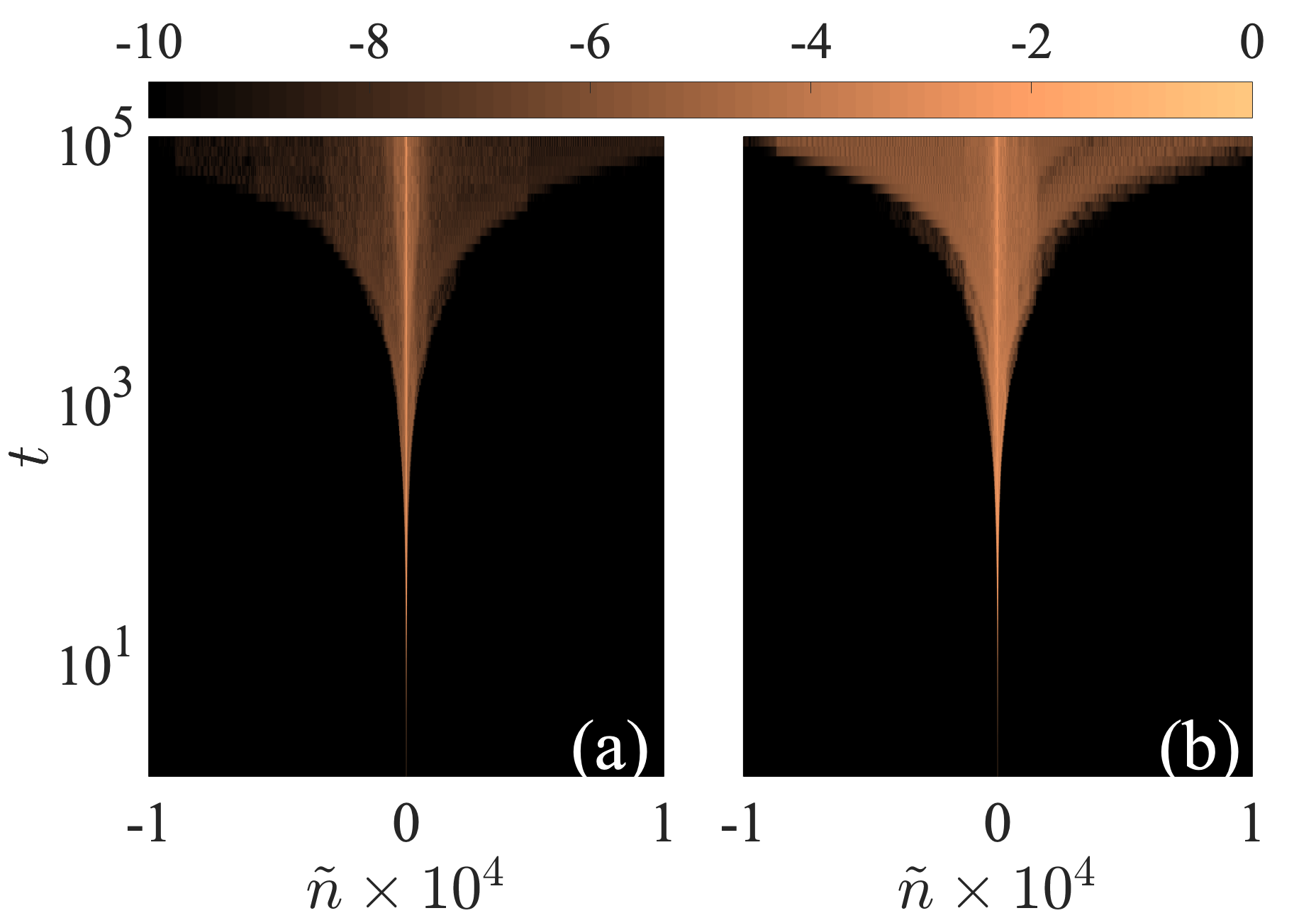}
\includegraphics[width=8.4cm]{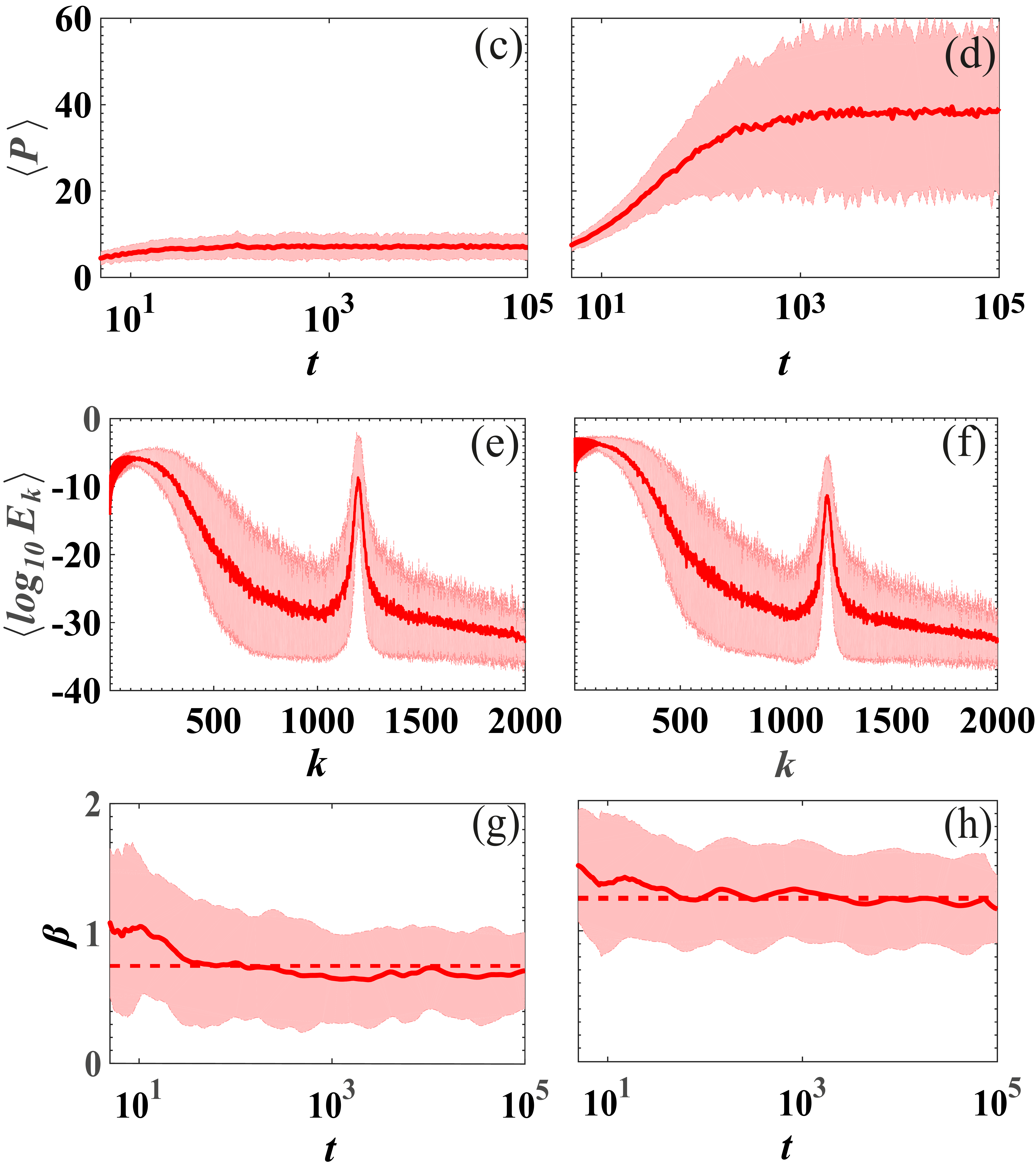}
\caption{Results corresponding to rotational (left panels) and transverse (right panels) initial momentum excitation. (a)-(b) Time evolution of the energy distribution for a representative disordered realization with colorbar in $\log_{10}$ scale . The horizontal axis represents $\tilde{n} = n -N/2$. (c)-(d) Time evolution of the average participation number $\langle P \rangle$. (e)-(f) Average energy per mode after projecting the initial condition to the normal modes. (g)-(h) Estimation of the exponent $\beta$ related to the time evolution of the average second moment through $\langle m_2(t) \rangle \propto t^{\beta}$. The horizontal dashed line indicates the values (g) $\beta=0.75$ and (h) $\beta=1.25$. For (c)-(h), results have been averaged over $200$ disorder realizations, and the shaded area denotes one standard deviation.}
\label{fig3}
\end{figure}
We first study the dynamics of the lattice under two different initial momentum excitations
\begin{align}
P_{N/2}^{(\phi)}(0)=\sqrt{I},\;{\rm or } \; P_{N/2}^{(u)}(0)=1
 \end{align}
i.e.,~initially exciting either the transverse or the rotational momentum of the central site. Note that this choice of initial conditions corresponds to
a total energy of $H=0.5$ for both cases.
Some typical time evolutions of the energy densities are shown Figs.~\ref{fig3}(a) and (b) for an initial (a) rotational and (b) transverse momentum excitation. We observe that for both cases, a large amount of energy remains localized in the region of the initial excitation at the lattice's center. This is expected due to the fact that most of the modes are localized and thus the implemented initial condition strongly excites localized modes around the central site. This is also quantified by the time
evolution of the averaged participation number $\langle P\rangle$ shown in Figs.~\ref{fig3}(c) and (d) for respectively the rotational and transverse initial momentum excitations. In fact we observe that in both cases $\langle P\rangle$ saturates to a small number compared to the total lattice size. However, comparing the two final values we observe a significant difference between the two cases as
the transverse initial excitation [Fig.~\ref{fig3}(d)] leads to a larger $\langle P \rangle$.

This behavior can be understood by studying the projection of the initial conditions onto the normal modes of a finite but large disordered lattice.
For a given initial momentum excitation we define the vector $\vec{V}(0)=[\dot{u}_1(0),\ldots, \dot{u}_N(0),\dot{\phi}_1(0),\ldots, \dot{\phi}_N(0)]^T$ whose projection on the system's normal modes is given by $\vec{\dot{R}}=\mathbf{A}^{-1}\vec{V}(0)$ with matrix $\mathbf{A}$ having as columns the lattice eigenvectors. Using this projection, we can calculate the energy given to each normal mode as $E_k=
\dot{R}_k^2/2$ where $\dot{ R}_k $ are the elements of projection vector $\vec{\dot{R}}$. Obviously the system's total energy is $H= \sum E_k$. Figs.~\ref{fig3}(e) and (f), although they appear to have a similar form, exhibit important differences regarding low index ($k$) modes (see also Appendix~\ref{sec7}). Since we sorted the modes with increasing frequency, low indices correspond to low frequency extended modes. In fact, for the initial 
\textit{rotational} momentum~[Fig.~\ref{fig3}(e)], the low frequency extended modes (low index $k$) are the stronger excited ones with an energy up to the order of  $10^{-6}$. On the other hand, by initially exciting the transverse momentum, the low frequency modes (low index $k$) as shown in Fig.~\ref{fig3}(f), are strongly excited acquiring energies up to $10^{-3}$. These orders of magnitudes difference in energy of low frequency extended modes explains the differences in $\langle P\rangle$ shown in Figs.~\ref{fig3}(c) and (d).

Here we also observe major differences between the micropolar lattice and the well studied $1$D harmonic lattice with disorder~
\cite{disorder_15,Kundu,Lepri}. With an initial momentum excitation, instead of exciting all modes with the same energy as in the $1$D harmonic lattice, here we observe a strong excitation of the low frequency modes and another set of modes around the cut-off frequency of the upper branch of the periodic case. As we will see below this has consequences to the energy transport.

To quantify the energy spreading we compute the averaged second moment $\langle m_2 \rangle $~\cite{Kundu,flach2009,SGF13,Bob2018,malcolmDNA,arn_gran} of the energy distributions, which for an initial excitation in the middle of the lattice is given by $m_{2} =  \sum_n (n-N/2)^2 h_n /H$.
Assuming a polynomial dependence of the spreading, for sufficiently long times, we may write $\langle m_2 \rangle  \propto t^{\beta}$ and the parameter 
$\beta$ is used to quantify the asymptotic behavior. The exponent $\beta$ is calculated by first smoothing the $m_2 (t)$ values of each disorder realization through a locally weighted difference algorithm~\cite{cleveland1,cleveland2}. The estimate of the rate of change
\begin{equation}
\beta=  \frac{ d  \log_{10}\langle m_2 (t) \rangle }{ d  \log_{10} t},
\end{equation}
is thus obtained numerically through a central finite difference scheme as the values of $m_2 (t)$ are analyzed in log-log scale. 

In Figs.~\ref{fig3}(g) and (h) we observe that for both cases $\beta$ reaches an asymptotic value. In fact $\beta \approx 0.75 ~(\beta \approx 1.25)$ for initial rotational (transverse) momentum excitations corresponding to subdiffusive (superdiffusive) transport. These values are quite different than the ones observed for the $1$D harmonic lattice where momentum excitation is always found to be superdiffusive with $\beta \approx 1.5$~\cite{Kundu,Lepri,Wagner}. To qualitatively explain this difference we first note that the exponent $\beta$ has been shown to depend mainly on two factors: (i) the characteristics of extended modes (group velocity, localization length as function of frequency, total number) and (ii) the projection of the initial condition on the modes~\cite{Kundu,Lepri,Wagner}. 
Regarding point (i), for both models, there is a set of extended modes at $\Omega \ll 1$. Major differences are thus expected since the dispersion relation of Eq.~(\ref{dispr}) for the micropolar lattice at low frequencies is quadratic with respect to the wavenumber i.e.,~ $\Omega \approx 3\sqrt{K^{(2)}}Q^2$ in contrast to the $1$D harmonic lattice where $\Omega \approx Q$. 
Furthermore, for the micropolar lattice, the \textit{quasi-extended} modes at higher frequencies may influence the energy spreading, as it was shown
for example in~\cite{Kundu,Luck} where additional extended modes were found either due to symmetries or resonances.
As far as point (ii) is concerned, the results of Figs.~\ref{fig3}(e) and (f) are substantially different from those of the $1$D harmonic lattice
indicating that differences between the two models are anticipated.

\subsection{Displacement excitation} \label{ssec2}
\begin{figure}
\includegraphics[width=8.5cm]{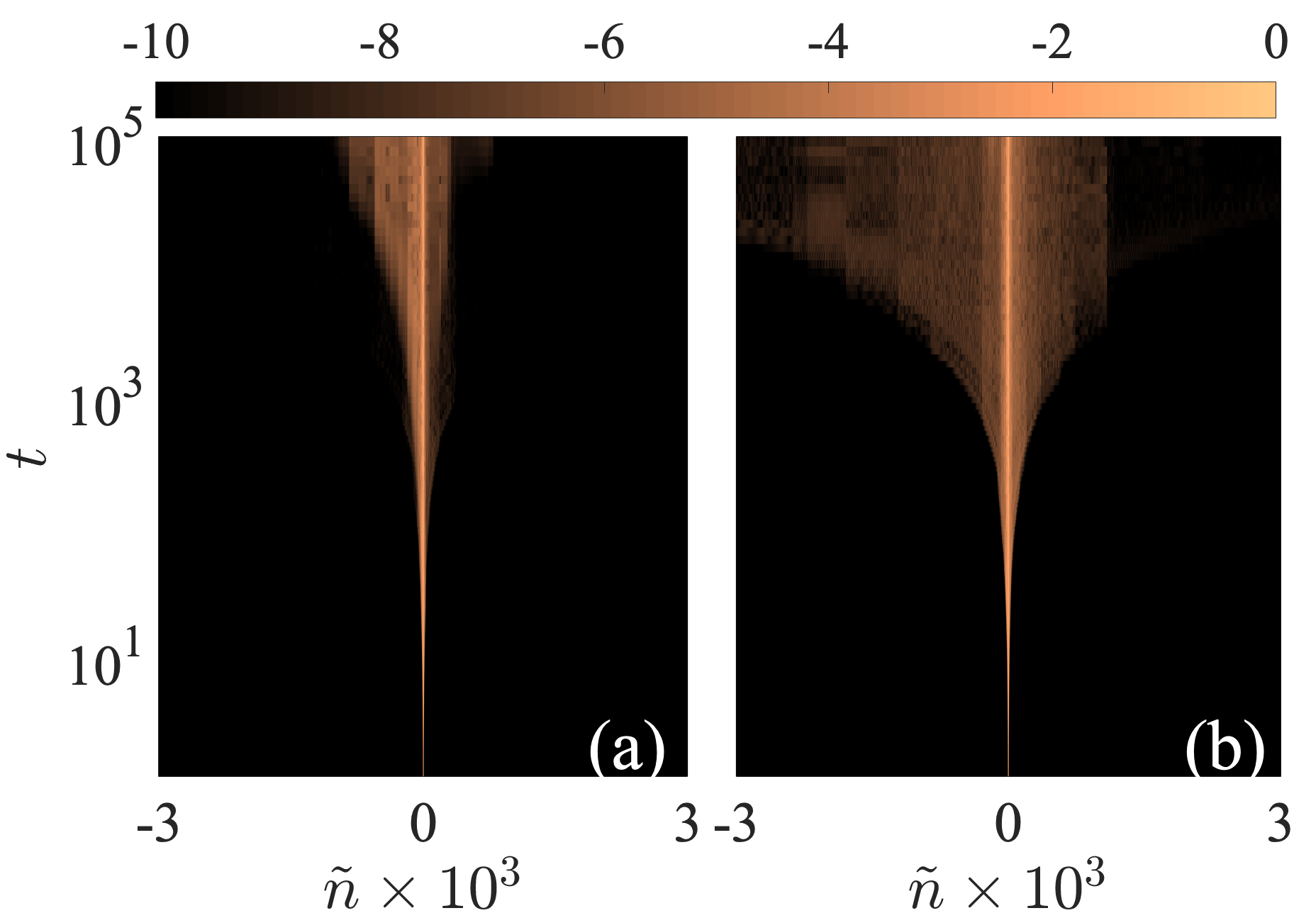}
\includegraphics[width=8.5cm]{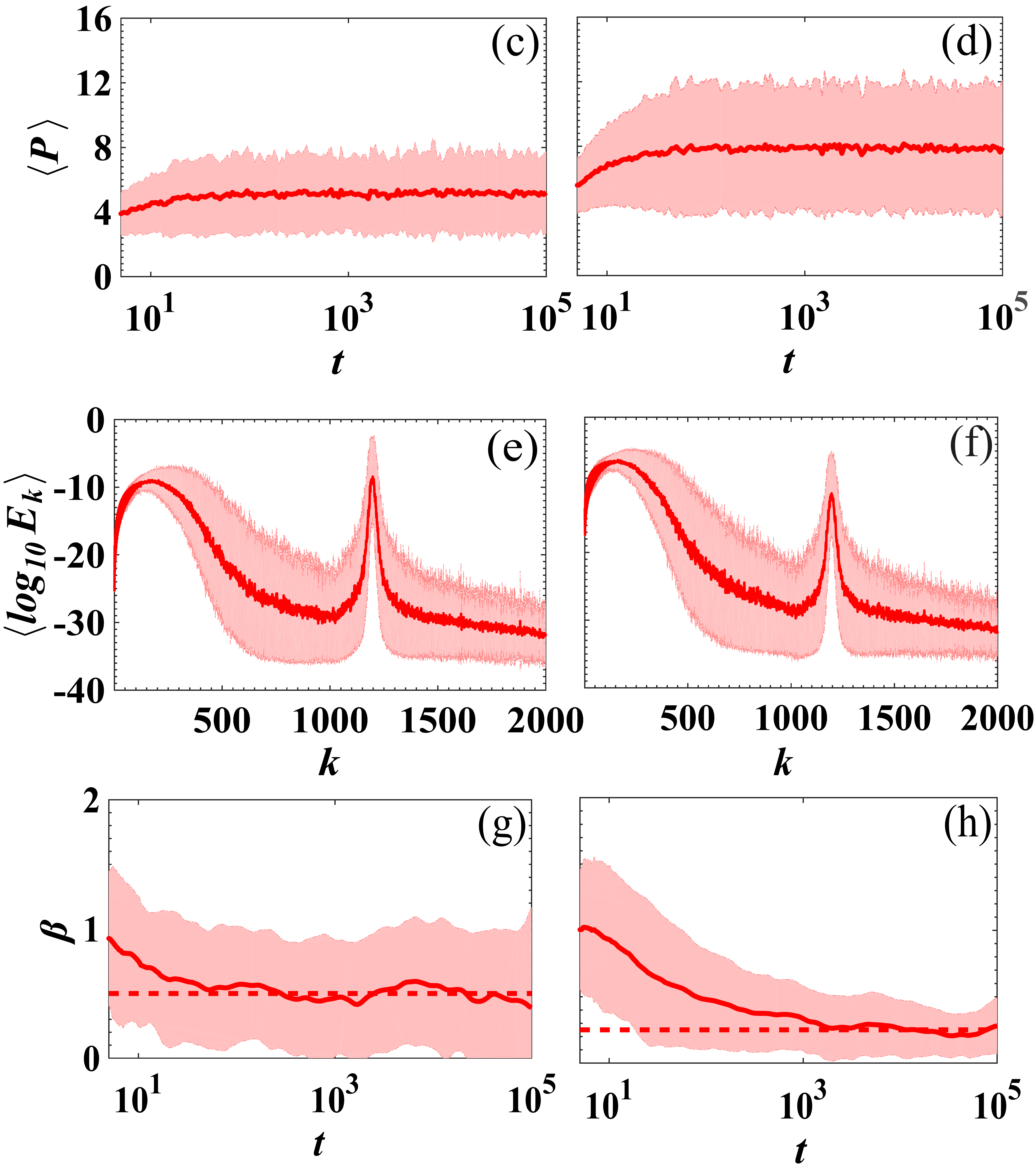}
\caption{ Similar to Fig.~\ref{fig3} but for displacement excitation(s). The horizontal dashed lines in (g) and (h) respectively indicate $\beta = 0.5$ and $\beta = 0.25$. }
\label{fig4}
\end{figure}
To further compare the behavior of the micropolar model to that of the $1$D harmonic lattice~\cite{Ishii,Kundu}, we now study the dynamics induced by the following initial conditions
\begin{align}
\phi_{\frac{N}{2}} (0)=\phi_{\frac{N}{2}},\;{\rm or } \; u_{\frac{N}{2}}(0)=u_{\frac{N}{2}},
 \end{align}
which correspond to initial rotation or transverse displacement of the central block. In this study, the values of $\phi_{\frac{N}{2}}$ and $u_{\frac{N}{2}}$ are chosen such that the total energy for each realization is again $H=0.5$.

Similarly to Section~\ref{ssec1}, the evolution of the energy distribution [Figs.~\ref{fig4}(a) and (b)] is characterized by a localized wave-packet at the region of the initial excitation in center of the lattice, and by a portion which is propagating.
However, compared to the initial momenta excitations, here the energy carried away from the central site is substantially smaller. For both types of initial conditions, $\langle P\rangle$ attains an asymptotic value of less than $10$ sites as shown in Figs.~\ref{fig4}(c) and (d). This behavior can be also understood using the projection of the initial conditions to the normal modes of a large but finite lattice. This is now done by projecting the vector $\vec{U}(0)=[u_1(0),\ldots, u_N(0),\phi_1(0),\ldots,\phi_N(0)]^T$  onto the normal modes to yield $\vec{R}=\mathbf{A}^{-1}\vec{U}(0)$. In this case, the energy of the $k$th normal mode is $ E_k = \Omega_k^2R_k^2/2$ with $\Omega_k$ being the $k$th eigenfrequency.
Again the system's total energy is $ H=\sum E_k $. The outcome of this projection is shown in Figs.~\ref{fig4}(e) and (f) for the initial rotation and transverse displacement respectively. The results are similar to those of Figs.~\ref{fig3}(e) and (f) with suppressed contributions of the low frequency modes leading to a small value of $\langle P\rangle$ during the evolution.

Furthermore, we also calculated the exponent $\beta$ for the energy propagation resulting from these two different initial excitations
and the results are shown in Figs.~\ref{fig4}(g) and (h). In a similar manner as in the $1$D disordered harmonic lattice case, single site displacement excitation lead to subdiffusive behavior. However, our findings show that although the initial rotation excitations leads to the same value ($\beta \approx 0.5$) as in the $1$D harmonic lattice, the initial transverse displacement features extremely slow energy transport with $\beta\approx 0.25$.
The discrepancy between the two models is anticipated  as it was discussed at the end of Section~\ref{ssec1}. However our results for the displacement excitations of the micropolar lattice strongly suggest that the energy transport is indeed mediated by both the low frequency and the \textit{quasi-extended} modes. To be more precise we compare the results of Fig.~\ref{fig3}(e) with Fig.~\ref{fig4}(e) and notice that in the latter
lower frequency modes are less excited leading to a smaller value of $\beta$ ($\beta\approx 0.75$ for the former and $\beta\approx 0.5$ for the latter). In the same spirit, by comparing Fig.~\ref{fig3}(e) with Fig.~\ref{fig4}(f) the main difference lies in the \textit{quasi-extended} modes which are suppressed in the later case leading to a $\beta\approx 0.25$ instead of $0.75$. Thus reducing the amount of energy allocated to either the low frequency extended modes or the \textit{quasi-extended} modes results in a reduced $\beta$ suggesting that both contribute to the energy spreading.

Note that the result of initial transverse momentum, corresponding to Fig.~\ref{fig3}(f) is not compared with the other three since in that case the low frequency modes are highly excited. 
We thus conclude that the complete picture is comprehended by casting one eye on the low frequency extended and the other onto the \textit{quasi-extended} ones.

\subsection{Energy contributions in the micropolar } \label{ssec4}
The participation number measures the localization of the total energy and the exponent $\beta$ is a measure of how fast the energy is spreading,
however, none of them carries any information of what amount of this energy
is attributed to the rotational or the transverse DOFs.
We can have an indication of how much energy is attributed to each of the two
DOFs by decomposing the total energy of the system into two parts i.e.,~$H =  H_{R} + H_{T}$ as follows
\begin{widetext}
\begin{align}
  &H_{R}   = \sum_{n=1}^{N}   \frac{1}{2I}  {P^{(\phi)}_{n}}^{2} 
+   \frac{ 9 }{8} K^{(1)}_{n+1}  (  \phi_{n+1} +  \phi_{n}   ) ^2 + 
  \frac{1 }{2} K^{(2)}_{n+1} (   \phi_{n+1} -    \phi_{n}  ) ^2 +\frac{3 }{4} K^{(1)}_{n+1} ({ \phi_{n+1} }  { }+ {  \phi_{n} }  )  (  { u_{n+1}}  -  { u_{n}} ),
\label{ham3} \\
 & H_{T}  = \sum_{n=1}^{N}    \frac{1}{2} {P^{(u)}_{n}}^{2}+   \frac{1 }{2} K^{(1)}_{n+1} (u_{n+1}   - u_{n})^2 +
\frac{3  }{4} K^{(1)}_{n+1} ({ \phi_{n+1} }  { }+ {  \phi_{n} }  )  (  { u_{n+1}}  -  { u_{n}} ),
\label{ham4}
\end{align}
\end{widetext}

\begin{figure}
\includegraphics[width=8.00cm]{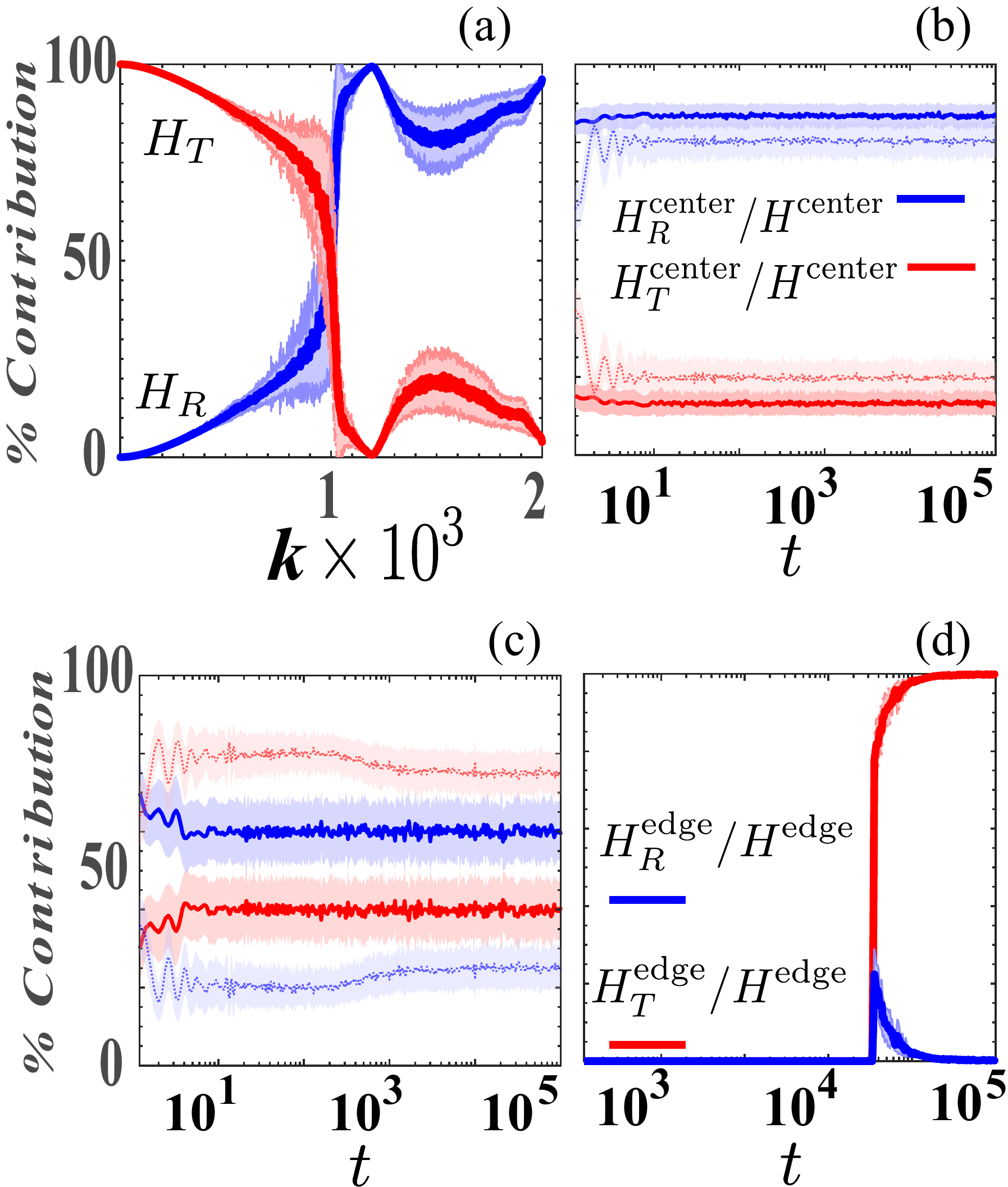}
\caption{ (a) Averaged normalized energy contributions $H_{R}$ and $H_{T}$ of the normal modes for finite lattices of $1000$ sites.
(b) Time evolution of normalized averaged rotational ($H_R^{center}$) and transverse ($H_T^{center}$) energy contributions near the excitation region. Solid bolder (dashed lighter shaded) curves show rotational displacement (momentum) initial excitations. (c) Same as (b) but for transverse displacement (momentum) initial excitations. (d) Time evolution of averaged normalized energy contributions $H_R^{edge}$ and $H_T^{edge}$, for transverse momentum excitations. Averaged values are over $200$ disorder realizations and one standard deviation is indicated by the lightly shaded regions.}
\label{fig6}
\end{figure}
\noindent separating the rotational $H_{R} $ and transverse $H_{T}$ energy contributions.
Note that the coupling potential energy,
which is described by the last terms in both Eqs.~(\ref{ham3}) and (\ref{ham4}), is equally shared between the two contributions.
It is interesting to determine the nature of the lattice's energy in two different regions: i) around the initially excited central block and ii) sufficiently far away from the region of localization. For the central area we calculate the energy using Eqs.~(\ref{ham3}) and (\ref{ham4}) but taking the sum for $n \in [N/2 -100,N/2+100]$ to obtain $H_R^{center}$ and $H_T^{center}$. Conversely, we also define the  energies at the edges of the energy distribution $H_R^{edge}$ and $H_T^{edge}$ by summing Eqs.~(\ref{ham3}) and (\ref{ham4}) for $n\notin [N/2-5000,N/2+5000]$. 
 
Before discussing the dynamical behavior of the system in these two distinct regions, it is relevant to show how the two different energy contributions are shared between the modes of a finite disordered lattice. The result is shown in Fig.~\ref{fig6}(a) where the red (blue) curve depicts the transverse (rotational) energy contribution. As a general observation, we mention that the modes with lower $k$ values are dominated by transverse motion, while the high frequency ones are dominated by rotational motion. As it is shown in Fig.~\ref{fig6}(b), for both initial conditions concerning the rotational DOFs [$P_{n}^{(\phi)}(0)=\sqrt{I}\delta_{n,N/2}$ and $\phi_{n} (0)=\phi_{N/2} \delta_{n,N/2}$], the central, localized part of the energy distribution is dominated by the rotational motion with almost the same ratios. 
 This is so as the majority of the localized modes are dominated by rotation [see Fig.~\ref{fig6}(a) and Fig.~\ref{profile}(b)].  
By initially exciting the transverse DOF we end up with the two energy contributions in the central part shown in Fig.~\ref{fig6}(c). We find that the central part of the energy distribution for the initial displacement excitation [$u_{n} (0)=u_{N/2} \delta_{n,N/2}$] is still dominated by rotation as indicated by the solid curve in Fig.~\ref{fig6}(c). Interestingly, for the case of initial transverse momentum excitation [$P_{n}^{(u)}(0)=\delta_{n,N/2}$], the energy contribution of each motion in the central part is inverted with respect to all other cases. This is due to the fact that this initial condition excites more strongly the low frequency modes [Fig.~\ref{fig3}(f)] which according to Fig.~\ref{fig6}(a) are dominated by transverse motion.

Let us now turn our attention to the energy distribution far away from the central site following the propagating tails that are responsible for the energy 
transfer. The corresponding results for the initial transverse momentum excitation is shown in Fig.~\ref{fig6}(d). After the arrival of the propagating front at the chosen sites $n=N/2\pm 5000$, it is readily seen that the energy at the edges is completely carried by the transverse motion. We have confirmed the same quantitative result for all types of initial conditions. This behavior can be comprehended since we have shown that energy is carried away mostly from the low frequency extended modes hence as shown in Fig.~\ref{fig6}(a) these modes (corresponding to small $k$) are almost completely constituted by transverse motion and so is the energy at the edges.

\subsection{The special case of $K^{(2)}=0$ } \label{ssec5}

Now, we focus on a special case of the system i.e.,~in the limit of vanishing bending stiffness
$K^{(2)}$. Note that such a case is very relevant to situations where the bending stiffness is so small that it may be neglected, see for example Ref.~\cite{Flornew}.Then, the corresponding dispersion relation of the periodic system [dashed lines in Fig.~\ref{profile}(a)] is substantially altered. In particular instead of two propagating bands, it consists of a zero frequency non-propagative band and a dispersive band emerging after a cut-off frequency $\Omega=2$. The zero frequency branch is made possible due to a counterbalance between the shear and bending forces~\cite{pichard}.
 
\begin{figure}
\includegraphics[scale=0.14]{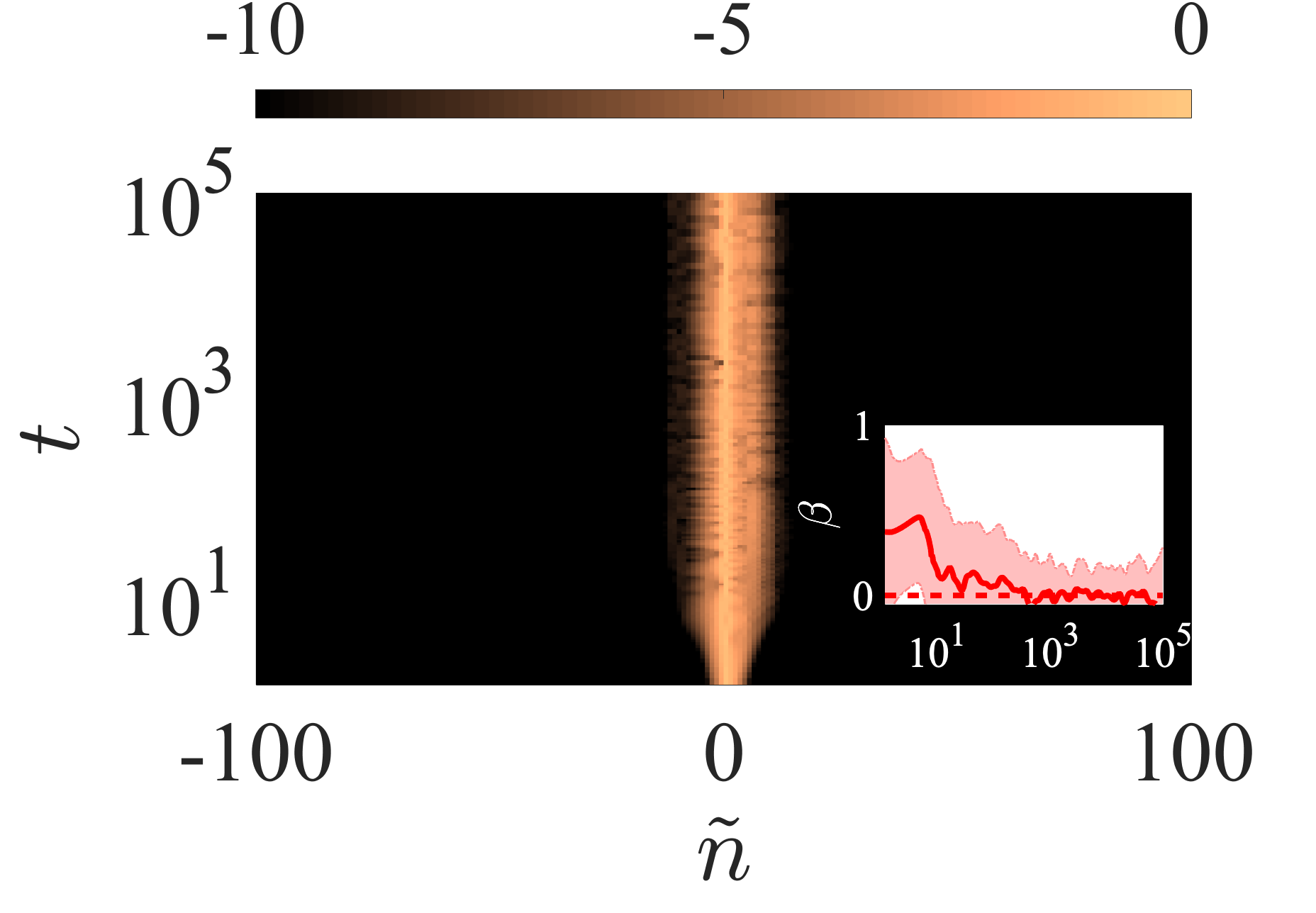}
\caption{ Time evolution of the energy density after an initial transverse momentum excitation $P_{N/2}^{u}(0)=1$ with  $K^{(2)}=0$ where $\tilde{n}=n-N/2$. The inset depicts the evolution of the exponent, $\beta$ in the relation $\langle m_2(t) \rangle \propto t^{\beta}$ averaging over $200$ disorder realizations, which is shown to be zero indicating no spreading. The horizontal dashed line indicates $\beta = 0$.}
\label{fig7}
\end{figure}

To study the energy transfer for this special case of $K^{(2)}=0$, we have performed simulations for the different single site initial conditions considered before and a characteristic example is shown in Fig.~\ref{fig7}. There it is readily seen that the energy remains localized around the center, and in addition there is no energy transfer to the rest of the lattice. This is also confirmed by the time dependence of the exponent $\beta$ in $\langle m_2(t) \rangle \propto t^{\beta}$, which becomes zero (see inset of Fig.~\ref{fig7}) thus signaling no energy spreading. 

Trying to explain the absence of energy transport we found (by solving the corresponding eigenvalue problem numerically) that the lower branch of the system's frequency spectrum, in the limit case of $K^{(2)}=0$, still remains at zero frequency even in the presence of 
strong disorder due to a counterbalance of the transverse and rotational motions. As such, in this limit, the micropolar lattice is similar to a $1$D KG model i.e.,~featuring a single propagating band emerging after a lower cut-off frequency and  thus the system is expected to exhibit AL.

\section{Conclusions} \label{sec4}

We have demonstrated how energy is transported in a strongly disordered micropolar lattice subject to shear forces and bending moments
when the shear stiffness are chosen randomly.
The phononic crystal investigated was composed of connected blocks possessing two degrees of freedom 
corresponding to transverse and rotational motion.
The dynamics of the energy density, under different single site initial excitations was characterized into two different regions: a localized energy 
distribution around the initially excited site and a propagating part at the edges of the lattice. The energy localization for each initial condition as quantified by the participation number $P$ was found to acquire a small (compared to the lattice length) asymptotic value.

Depending on which motion or momentum we initially excited, energy spreading was found to be either superdiffusive or subdiffusive,
as quantified by the energy's second moment $m_2$. Compared to the underlying $1$D harmonic case, energy transport is altered, and in general 
the micropolar lattice featured slower spreading. The modified energy transport characteristics are attributed to the differences 
of the dispersion relation between the two models in the low frequency limit, to the weight by which the modes of the system are excited depending
on the initial condition and also to the existence of additional \textit{quasi-extended} modes in the micropolar lattice.

Furthermore, by measuring  the parts of the total energy related to
the rotational and transverse motions we showed that the propagating part is always carried by 
translation for any choice of initial condition. On the other hand, the localized part was found to be either dominated by rotation or translation depending on the 
initial conditions. Finally, the limiting case of vanishing bending force was found to be similar to a linear $1$D KG lattice which exhibits AL and thus no energy spreading. 

Our results not only revealed interesting properties of $1$D disordered micropolar lattices with bending forces, but also raised new questions for future investigations. A direct generalization of our results is to study the effect of other kinds of disorder i.e.,~disorder in the masses or in different combinations of the stiffnesses. Furthermore, the appearance of the extended modes at the edge of the upper band is worthy of its own investigation in relation to other known models where anomalous localization appears either due to correlations or symmetry. 

\section*{acknowledgments}
Ch.~S.~thanks LAUM for its hospitality during his visits when part of this work was carried out. We also thank the center for High Performance Computing (\url{https://www.chpc.ac.za}) for providing computational resources for performing significant parts  of this paper's computations.
\appendix
\section{Quasi-extended modes} \label{sec6}
\begin{figure}[ht]
   \centering
     \includegraphics[scale=0.25]{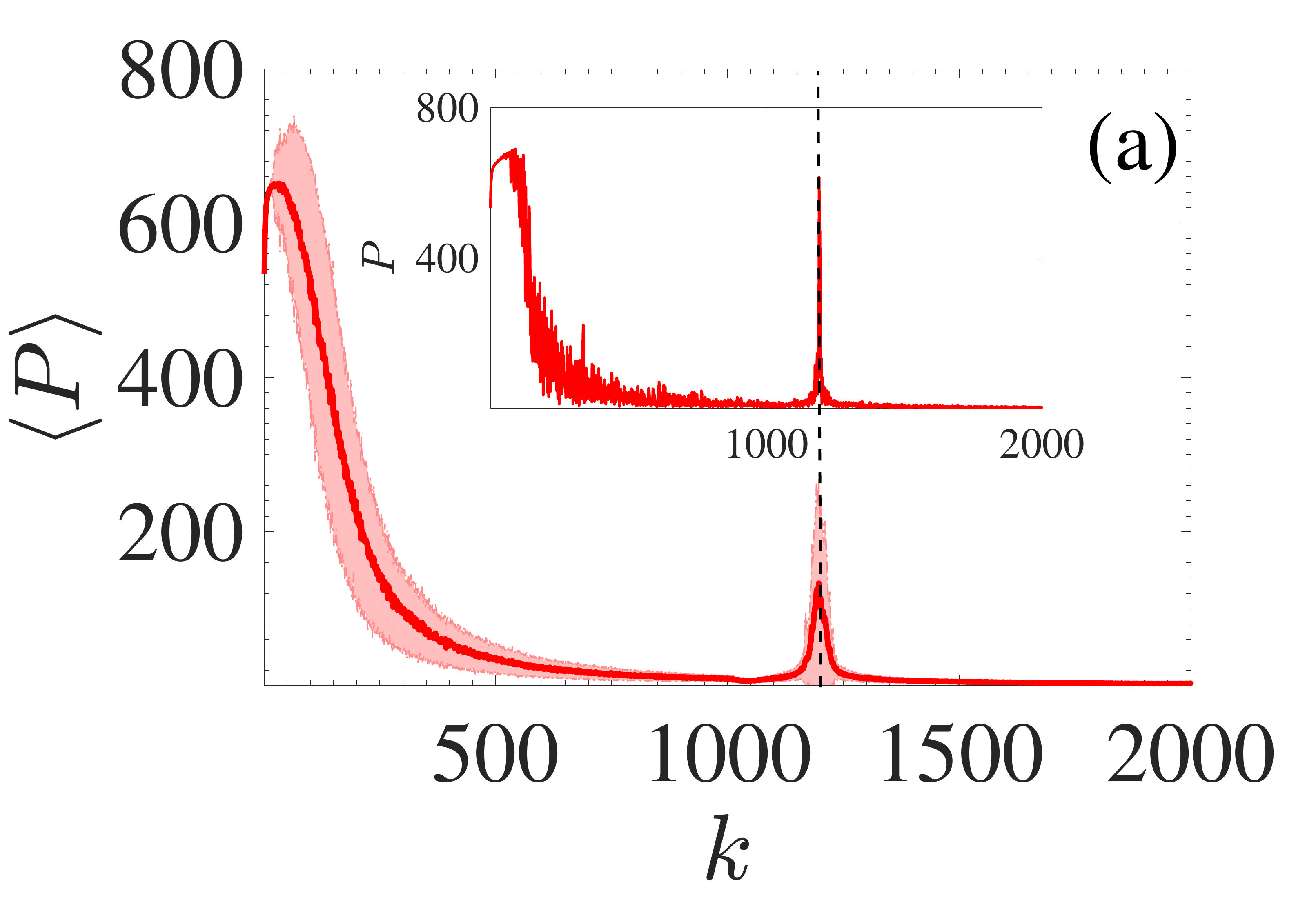} 
       \includegraphics[scale=0.25]{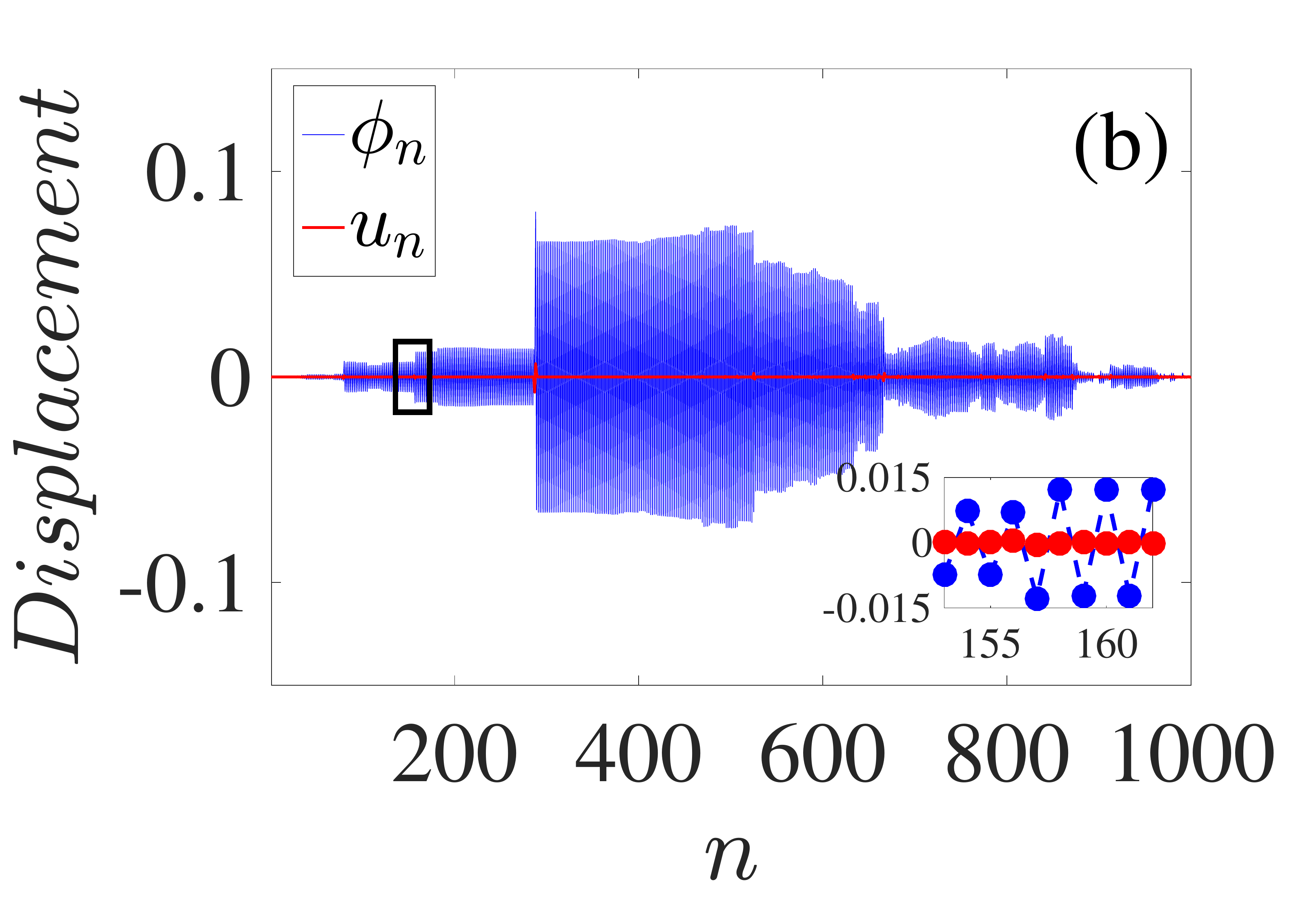} 
           \caption{ (a) A full view of the inset depicted in Fig.~\ref{profile}(b) and its own inset shows $P$ against sorted linear eigenmodes for a single disorder realization. The vertical dashed line denotes the index where the \textit{quasi-extended} modes appear.
             (b) The profile of a characteristic \textit{quasi-extended} mode $k=1194$ showing negligible displacements $u_n$ in comparison to $\phi_n$. The inset shows a zoom of the region enclosed by a black rectangle with consecutive rotations having similar amplitudes and opposite signs ($\phi_{n+1}\approx- \phi_n$). }
   \label{fig1a}
\end{figure}
Here we focus our attention on the \textit{quasi-extended} modes appearing close to the cut-off frequency of the upper band of the periodic case (see Fig.~\ref{profile}).
As depicted in Fig.~\ref{fig1a}(a), the participation number $\langle P \rangle$ features a peak around $k \approx 1200$ which corresponds to the cut-off frequency of the upper branch of the periodic system [see Figs.~\ref{profile}(a) and (b)].
Note that in many cases we found that these modes may be as extended and have $P$ values which are of the same order as the low index modes (small $k$) as indicated by the inset in Fig.~\ref{fig1a}(a) corresponding 
to a single realization. 

To further understand this phenomenon we now consider a characteristic profile of such a mode depicted in Fig.~\ref{fig1a}(b). We find that: (i) these modes consist almost solely of rotational motion (the contribution of the transverse DOFs is negligible i.e.,~$u_n \approx 0$ ) and (ii) the profile of the modes consists of various regions with consecutive rotations of similar amplitude and opposite signs ($\phi_{n+1}\approx- \phi_n$), as shown by the zoom in the inset of Fig.~\ref{fig1a}(b). Using this two observations and the functional form of the Hamiltonian~(\ref{hamilt}), it is noticeable that for these modes, effectively only the bending potential term analogous to $K^{(2)}$ is present. But since there is no disorder in $K^{(2)}$ these modes are extended reminiscent of the periodic lattice.
\section{Eigenmode Projections} \label{sec7}
\begin{figure}[ht]
   \centering
     \includegraphics[scale=0.25]{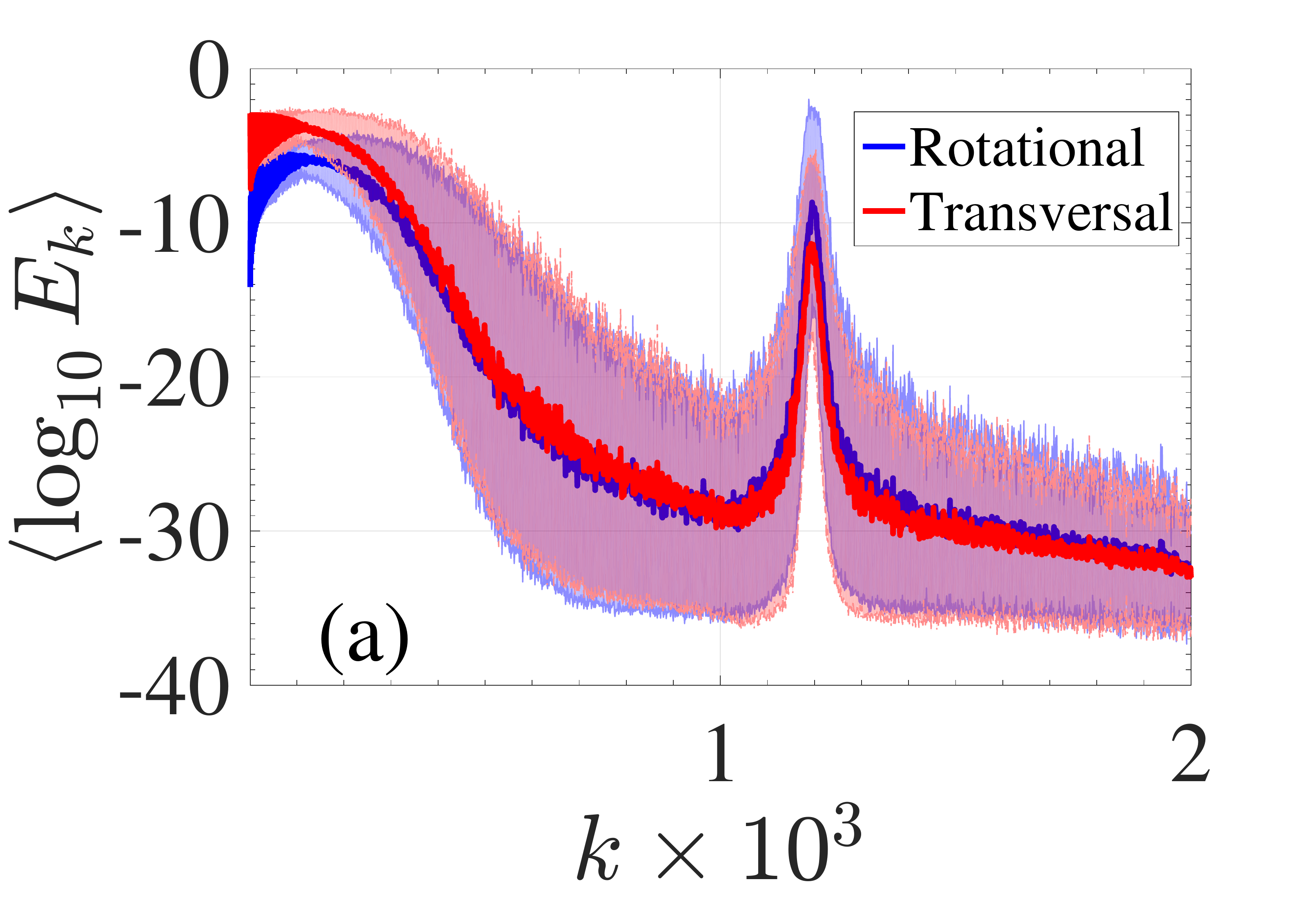} 
       \includegraphics[scale=0.25]{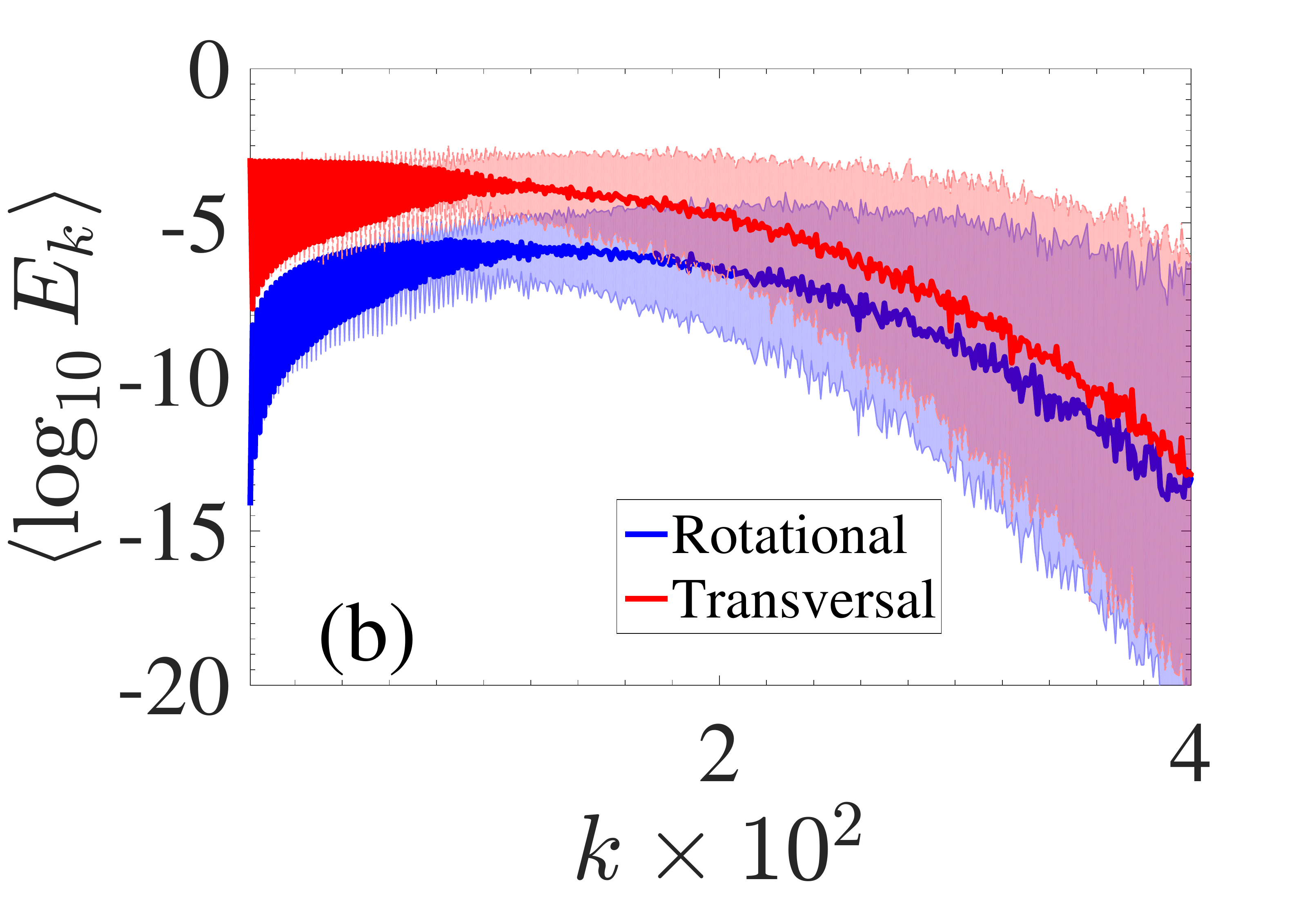} 
           \caption{ (a) Average (over $200$ disorder realizations) energy per mode after projecting the initial condition to the normal modes.
             (b) A zoom of panel (a) for small $k$. The lightly shaded red and blue regions indicate one standard deviation on either side of the mean value.}
   \label{fig2a} 
\end{figure}
\begin{figure}[ht]
   \centering
     \includegraphics[scale=0.25]{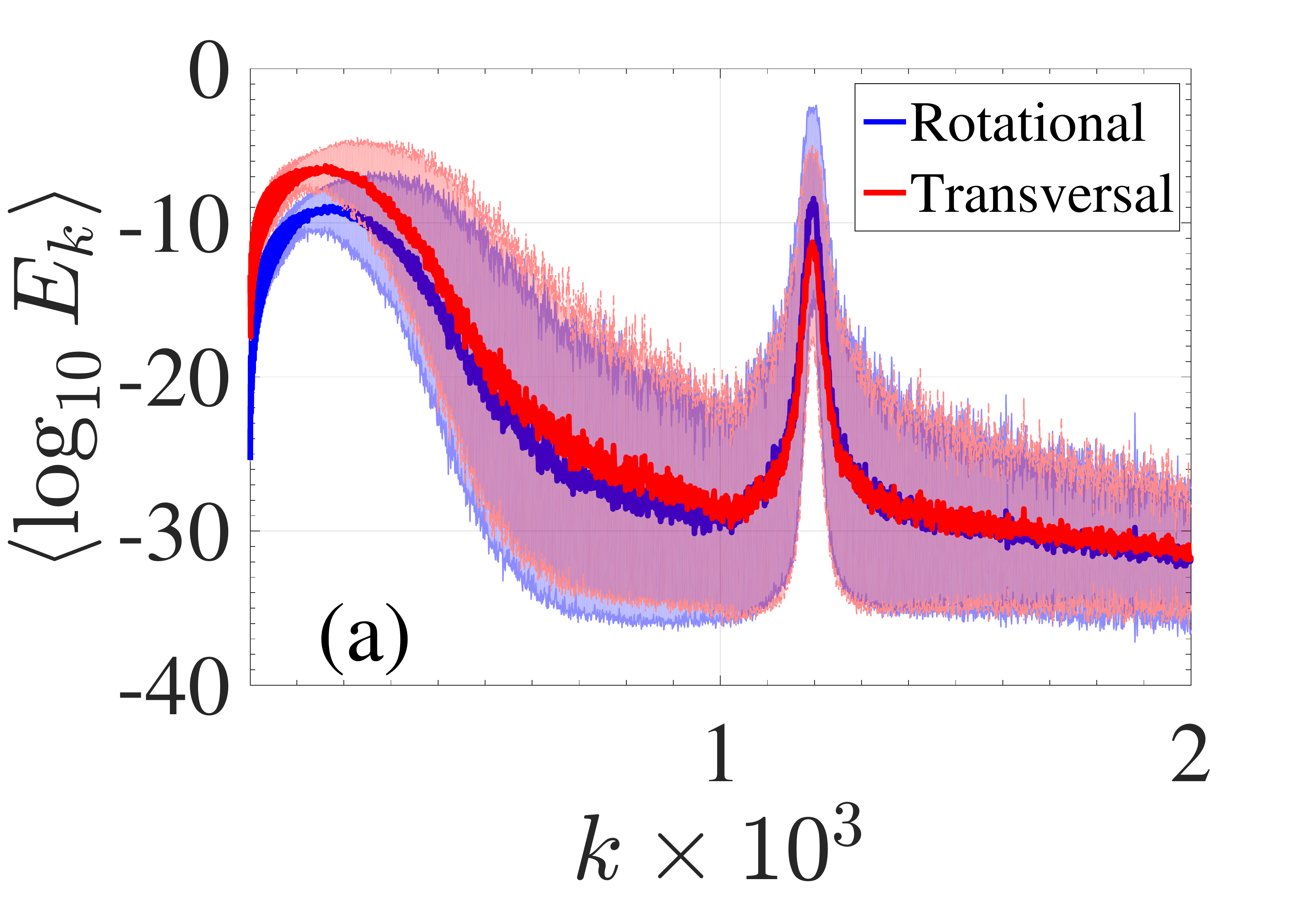} 
       \includegraphics[scale=0.25]{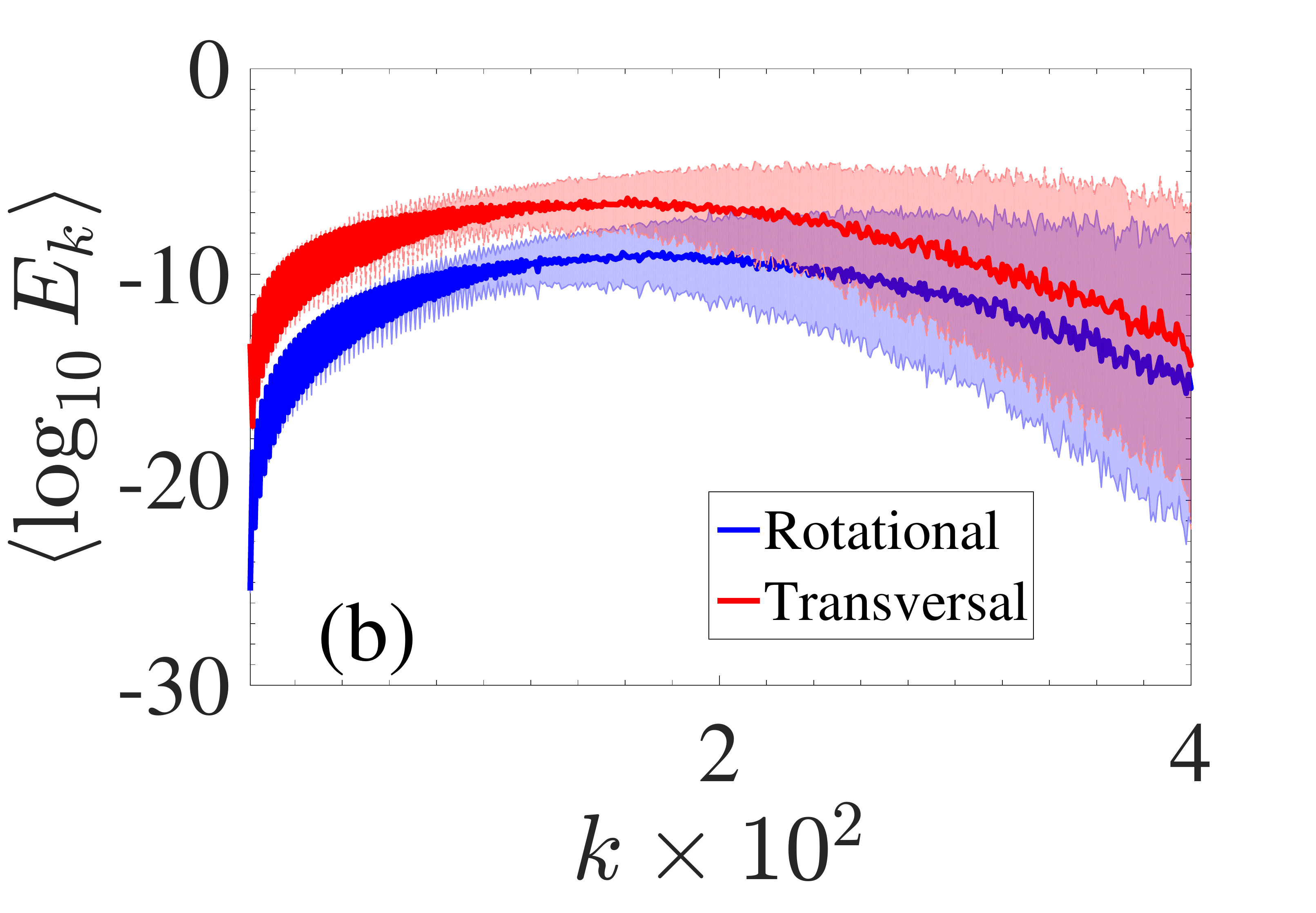} 
           \caption{ (a) Figs.~\ref{fig4}(e) and (b) combined together.
             (b) A zoom of some small $k$ values in (a). The lightly shaded red and blue regions indicate one standard deviation on either side of the mean value obtained over $200$ disorder realizations. }
   \label{fig3a}
\end{figure}

Here we take a closer look at the projections of the initial momentum excitations  onto the normal modes especially in the low frequency regime.
The results of Figs.~\ref{fig3}(e) and (b) are combined together for comparison in Fig.~\ref{fig2a}(a) while a zoom in the low frequencies is shown
in Fig.~\ref{fig2a}(b). From the latter we observe that for $k\rightarrow 0$ the energy difference between the two types of excitation is as much as $10$ orders of magnitude. This explains
the larger values of participation number $\langle P\rangle$ shown in Figs.~\ref{fig3}(c) and (d).
Similarly for the initial displacement excitations the results are shown in Fig.~\ref{fig3a}(a) and (b). Although here the two curves are more similar
again differences in the low frequency regime show that the initial transverse excitation (red curve) will acquire a larger $\langle P\rangle$ 
as it is found by comparing Figs.~\ref{fig4}(c) and (d).

\end{document}